\documentclass[12pt]{article}
\usepackage{graphicx}

\newcommand\pubnumber{SLAC--PUB--15858}
\newcommand\pubdate{December, 2013}

\def\SLAC{SLAC,
    Stanford University, Menlo Park, California 94025 USA}
\def\doeack{\footnote{Work supported by the US Department of Energy,
                     contract DE--AC02--76SF00515.}}

\def\Title#1{\begin{center} {\Large #1 } \end{center}}
\def\Author#1{\begin{center}{ \sc #1} \end{center}}
\def\Address#1{\begin{center}{ \it #1} \end{center}}

\newcommand\pubblock{\rightline{\begin{tabular}{l} \pubnumber\\
         \pubdate \end{tabular}}}
\newenvironment{Abstract}{\begin{quotation} \begin{center}
                       ABSTRACT
     \end{center}\bigskip  }{\end{quotation}}
\newenvironment{Presented}{\begin{quotation} \begin{center} 
             CONTRIBUTED TO\end{center}\bigskip 
      \begin{center}\begin{large}}{\end{large}\end{center} \end{quotation}}

\def\Acknowledgements{\bigskip  \bigskip \begin{center} \begin{large}
             \bf ACKNOWLEDGEMENTS \end{large}\end{center}}

\def\beq{\begin{equation}}
\def\eeq#1{\label{#1}\end{equation}}
\def\eeqn{\end{equation}}

\newenvironment{Eqnarray}%
   {\arraycolsep 0.14em\begin{eqnarray}}{\end{eqnarray}}
\def\beqa{\begin{Eqnarray}}
\def\eeqa#1{\label{#1}\end{Eqnarray}}
\def\eeqan{\end{Eqnarray}}

\def\leqn#1{(\ref{#1})}

\let\bar=\overbar

\def\etal{{\it et al.}}

\def\lsim{\mathrel{\raise.3ex\hbox{$<$\kern-.75em\lower1ex\hbox{$\sim$}}}}
\def\gsim{\mathrel{\raise.3ex\hbox{$>$\kern-.75em\lower1ex\hbox{$\sim$}}}}

\def\del{\partial}
\def\Dslash{\not{\hbox{\kern-4pt $D$}}}
\def\dslash{\not{\hbox{\kern-2pt $\del$}}}

\def\ee{e^+e^-}

\def\msb{{\bar{\scriptsize M \kern -1pt S}}}

\def\drb{{\bar{\scriptsize D \kern -1pt R}}}

\makeatletter
\def\section{\@startsection{section}{0}{\z@}{5.5ex plus .5ex minus
 1.5ex}{2.3ex plus .2ex}{\large\bf}}
\def\subsection{\@startsection{subsection}{1}{\z@}{3.5ex plus .5ex minus
 1.5ex}{1.3ex plus .2ex}{\normalsize\bf}}
\def\subsubsection{\@startsection{subsubsection}{2}{\z@}{-3.5ex plus
-1ex minus  -.2ex}{2.3ex plus .2ex}{\normalsize\sl}}

\renewcommand{\@makecaption}[2]{%
   \vskip 10pt
   \setbox\@tempboxa\hbox{\small #1: #2}
   \ifdim \wd\@tempboxa >\hsize     
       \small #1: #2\par          
     \else                        
       \hbox to\hsize{\hfil\box\@tempboxa\hfil}
   \fi}

 \def\citenum#1{{\def\@cite##1##2{##1}\cite{#1}}}
 
\newcount\@tempcntc
\def\@citex[#1]#2{\if@filesw\immediate\write\@auxout{\string\citation{#2}}\fi
  \@tempcnta\z@\@tempcntb\m@ne\def\@citea{}\@cite{\@for\@citeb:=#2\do
    {\@ifundefined
       {b@\@citeb}{\@citeo\@tempcntb\m@ne\@citea\def\@citea{,}{\bf ?}\@warning
       {Citation `\@citeb' on page \thepage \space undefined}}%
    {\setbox\z@\hbox{\global\@tempcntc0\csname b@\@citeb\endcsname\relax}%
     \ifnum\@tempcntc=\z@ \@citeo\@tempcntb\m@ne
       \@citea\def\@citea{,}\hbox{\csname b@\@citeb\endcsname}%
     \else
      \advance\@tempcntb\@ne
      \ifnum\@tempcntb=\@tempcntc
      \else\advance\@tempcntb\m@ne\@citeo
      \@tempcnta\@tempcntc\@tempcntb\@tempcntc\fi\fi}}\@citeo}{#1}}
\def\@citeo{\ifnum\@tempcnta>\@tempcntb\else\@citea\def\@citea{,}%
  \ifnum\@tempcnta=\@tempcntb\the\@tempcnta\else
  {\advance\@tempcnta\@ne\ifnum\@tempcnta=\@tempcntb \else\def\@citea{--}\fi
    \advance\@tempcnta\m@ne\the\@tempcnta\@citea\the\@tempcntb}\fi\fi}
\makeatother

\begin{document}
\begin{titlepage}
\pubblock

\vfill
\Title{Estimation of  LHC and ILC Capabilities for  Precision Higgs Boson Coupling Measurements}
\vfill
\Author{Michael E. Peskin\doeack}
\Address{\SLAC}
\vfill
\begin{Abstract}
This paper discusses some aspects of the estimates of Higgs boson
coupling sensitivities for LHC and  ILC presented at
the Snowmass 2013 meeting.  I estimate the measurement
accuracies underlying the CMS presentation to Snowmass. I present 
new fits for the ILC capabilities.  I present some joints fits to 
prospective LHC and ILC data that demonstrates the synergy of
the High-Luminosity LHC and ILC programs.
\end{Abstract}
\vfill
\begin{Presented}
Snowmass 2013 Electronic Proceedings \\
Community Summer Study, Minneapolis, MN \\
July 29 -- August 6, 2013\end{Presented}
\vfill

\newpage
\tableofcontents
\end{titlepage}

\def\thefootnote{\fnsymbol{footnote}}
\setcounter{footnote}{0}

\section{Introduction}

The evidence is accumulating that the resonance discovered
by ATLAS and CMS at 125~GeV~\cite{ATLAS,CMS} is a Higgs boson.
Both the intrinsic mystery surrounding this particle and the key
role of the Higgs sector in elementary particle physics make it 
an important goal to understand the properties of this particle
as accurately as possible.

The goals for the study of the couplings of the Higgs boson are 
discussed in detail in the Snowmass Higgs working group
report~\cite{Higgsworking}.   There are two sides to the story.
On one hand, the idea of the Standard Model that electroweak
symmetry is broken by a single complex doublet of scalar fields has
no compelling foundation.  It is just the simplest choice among a 
large number of options.  On the other hand, if there are additional
particles in addition to  the lightest Higgs boson but these
particles
are heavy, at mass $M$, the Decoupling Theorem~\cite{Haber} tells us that the
lightest
Higgs has the properties predicted by the Standard Model up to
corrections of order  $m_h^2/M^2$. At present, the properties of the 
125~GeV resonance agree with those of the Standard Model Higgs boson
to about 30\% accuracy.  This does not yet test the hypothesis of a 
single Higgs doublet.   To discover new structure in the Higgs sector,
we
need to look for effects at the 5\% level.  To discover such effects,
we 
need experiments that can explore the landscape of Higgs boson 
couplings in a model-independent way with accuracies at the 1\%
 level~\cite{Higgsworking}. 

One of the goals of the Snowmass 2013 study is to understand
the accuracy with which the couplings of the Higgs boson will be 
measured at the future stages of the LHC and at other future 
colliders.   An important question is the ultimate capability of the 
LHC experiments. The ATLAS and CMS
experiments provided estimates of this capability for 300~fb$^{-14}$
and 3000~fb$^{-1}$ data sets at 14~TeV in their White
Papers~\cite{ATLASWP,ATLASnew,CMSWP}.

The CMS White Paper gave estimates that were more optimistic 
and also took account of possible future improvements in our 
ability to extract information from the LHC data.  However, the
results of this paper were not presented in a way that makes it
straightforward to evaluate the capability of CMS to discover or
exclude specific theoretical models or, indeed, to carry out any 
fits other than those specifically included in the paper.

The first purpose of this note is to suggest a way to remedy this 
difficulty.  I provide a model of the CMS analysis in term of 
measurements accuracies for a list  of Higgs boson process
that can be considered as independent measurements.  
The second purpose of this note is to carry out fits 
involving the new estimates of ILC capabilities presented 
at Snowmass~\cite{ILCHiggs}.   With 
both sets of  inputs in hand, I then carry out some 
fits for the combined capabilities
 not yet included in the Snowmass documentation.

This paper is organized as follows:  In Section 2, I present my 
methodology for estimating Higgs boson coupling accuracies.
In Section 3, I present an interpretation of the CMS results in 
this framework.   In Section 4, I discuss the effect on the LHC
fits of including invisible modes of Higgs decay.
In Section 5,  I carry out a 10-parameter fit similar to that 
proposed in
\cite{myHiggsBR}  to quantify the accuracies of model-independent
ILC Higgs coupling determinations at the various ILC stages.
In Section 6, I present joint fits that make use of LHC and ILC
results.
In Section 7, I give some editorial comments.

The opinions expressed in this paper are strictly my own.  They should
not be mistaken for opinions of the Snowmass Higgs working group 
or opinions
of the Snowmass Energy Frontier conveners.

\section{Methodology}

In this paper, I will parametrize deviations of the couplings of the 
125~GeV Higgs boson as
\beq
  \kappa_A =   g(hA\bar A)/(SM) \ ,
\eeq{kappadef}
where $g(hA\bar A)$ is the coupling of the Higgs boson to the 
$A\bar A$ final state defined on the Higgs mass shell and, always
in this paper, $(SM)$ indicates the Standard Model expectation.
I will treat the Higgs boson in the narrow resonance approximation.
This is a very good approximation, since the Standard Model 
expectation for the width of the Higgs boson is about 4~MeV.

Couplings induced at the loop level in the Standard Model may 
receive contributions from the heavy particles of the Standard Model,
and also from new particles that are not 
yet known.  In this analysis, I will consider the couplings of the 
Higgs boson to $gg$ and $\gamma\gamma$ to be parametrized
by $\kappa$ values that are independent of those for $t$ and $W$,
which give the largest contributions to the purely Standard Model loop
effects.  In this paper, I will ignore the minor modes $h\to Z\gamma$
and $h\to \mu^+\mu^-$.  

Total cross sections and ratios of branching ratios have a simple
dependence on the $\kappa_A$, for example,
\beq
  \sigma(\ee \to Zh)/(SM) = \kappa_Z^2 \ , \qquad 
  {BR(h\to ZZ^*)/(SM)\over BR(h\to \gamma\gamma)/(SM)} =
  \kappa_Z^2/\kappa_\gamma^2 \ .
\eeq{kappaforsigma}
However, collider experiments more typically measure the rate for a 
complete process of Higgs production and decay to a particular final
state.   The ratio of this rate to the Standard Model expectation is
given by
\beq
  \sigma(A\bar A \to h) BR(h\to B\bar B )/(SM) =  {\kappa_A^2
    \kappa_B^2\over \kappa_h^2 }\ ,
\eeq{kappasigBR}
where $\kappa_h$ is the scale factor for the Higgs total width.
Within the Standard Model,
\beq
    \kappa_h^2 = \sum_C \kappa_C^2 BR(h\to C\bar C)|_{SM} \ .
\eeq{kappahdefprelim}

In this paper, I will also consider the possibility of decay modes
not included in the Standard Model.   These include {\it invisible} 
decays, for example, the decay of the Higgs boson to a pair of 
dark matter particles that cannot be seen by a collider detector.
Other exotic modes of Higgs decay, to visible light particles or to 
long-lived particles outside the Standard Model, are also possible.
In this context, it is important to distinguish {\it invisible}
decays---which can be measured at the LHC or the ILC in processes
 in which 
the Higgs is produced along with a $W$ or $Z$ boson or with jets
tagging vector boson fusion---from {\it unrecognizable} decays---for 
which there is no strategy for observing the Higgs decay mode above
background.  At the LHC, the decay $h\to c\bar c$ is {\it
  unrecognizable}.  It is 
easy to imagine exotic decays, for example, to multiple
jets,
 that also could not be 
discovered by the LHC experiments.   The decay $h\to gg$ cannot 
be observed at the LHC, but I will treat it as an observable decay, 
because the $h\to gg$ decay width is directly proportional to the $gg$
fusion production cross section and is thus directly constrained by LHC 
Higgs measurements.   The full formula for $\kappa_h^2$ is
\beq
    \kappa_h^2 = \biggl(\sum_C \kappa_C^2 BR(h\to C\bar
    C)|_{SM}\biggr)/(1 - BR_{inv} - BR_{exotic} - BR_{unr}) \ ,
\eeq{kappahdef}
where $C$ runs over observable Standard Model modes of Higgs
decay,  $BR_{inv}$ is the branching ratio to invisible modes, 
$BR_{exotic}$ is the branchig ratio to exotic, detectable modes, 
and $BR_{unr}$ is the branching ratio to unrecognizable modes.

The appearance of $\kappa_h$ in the formula \leqn{kappasigBR} 
couples all of the $\kappa_A$ into the   interpretation of any 
single rate measurement.  This the major difficulty  to be overcome 
in making a model-independent interpretation of the rate measurements
in terms of Higgs couplings.  Special difficulties arise at hadron
colliders,
because, in the Standard Model, the decay $h\to b\bar b$ account for
over 50\% of the total width and, at the same time, this decay is
exceptionally
difficult to observe above the hadron collider backgrounds.
   
This paper will be concerned with estimating the expected errors on
the  parameters $\kappa_A$ in future accelerator programs.   The
methodology 
of this paper will be very simple, even naive, but I hope that its 
transparency will useful for further investigations. I consider
a set of 10  parameters:
\beq
 \kappa_W\  , \  \kappa_Z \ ,\  \kappa_b \ ,\  \kappa_g \ ,\
 \kappa_\gamma \ , \ 
 \kappa_\tau \ , \ \kappa_c \ ,  \ \kappa_t \ ,  \ BR_{inv} \ ,
 \ BR_{unr} \ .
\eeq{totkappa}
Note that I assume for simplicity that the Higgs boson has no
detectable exotic decay modes.
For each fit, I give a prescription that defines the 10
parameters
in terms of an underlying set.  I specify a set of input
measurements
that constrain these underlying parameters.  Each measurement
is  considered to be a strictly independent piece of data, centered on 
the SM expectation with a
Gaussian error distribution.  The error is be determined by 
adding statistical and systematic errors in quadrature.  By multiplying
together the Gaussians and applying other needed constraints (for 
example, $BR_{inv} \geq 0$), I obtain a likelihood function for
the
underlying $\kappa_A$ values.  I explore this function using the 
VEGAS integrator~\cite{VEGAS}, histogram the relevant variables,
 and quote the 1 $\sigma$ confidence
intervals in each variable.

The 
systematic errors quoted include the theoretical errors necessary to
extract the rate (for example, the uncertainties in the 
SM expectation for the cross 
sections for the relevant production processes) but not the
uncertainty
 in the computation of the SM value of $g(hA\bar A)$. 
At the ILC, at least, the Higgs partial widths are extracted in 
a model-independent way, and then these values can be directly 
compared to SM calculations.  The uncertainties in 
the  SM values of Higgs
partial widths,
as they are currently 
quoted~\cite{HiggsBRs,HiggsBRstoo,WellsHiggsBRs},  are dominated by the
uncertainties in  input parameters such as $\alpha_s$ and $m_b$.
These are expected to improve greatly over the time scale relevant
to these projections~\cite{Mackenzie}.   It is worth emphasizing that
both experiment and theory must improve for the values of the Higgs
partial widths in order to make tests of these couplings with errors at the
1\% level.

\section{Interpretation of the CMS results}

In principle, it is straightforward to use the method 
described in the previous section to produce 
estimates of the Higgs coupling uncertainties from future collider
programs.
One must write down a list of input measurements, estimate the error for
each, construct the likelihood function described in the previous
section,
     and measure its properties.  Unfortunately, the presentations to 
Snowmass from the LHC collaborations \cite{ATLASWP,ATLASnew,CMSWP} do not 
provide the information needed for such an analysis. 

ATLAS 
has presented its estimates for its capabilities for Higgs measurements in a very 
explicit way, including as numerical tables.  
The uncertainties presented at Snowmass~\cite{ATLASWP} were quite 
conservative.  For example, the error on
the
rate for Higgs production and decay to $WW$ was projected not to change from 
300~fb$^{-1}$  to 3000~fb$^{-1}$.  More recently, ATLAS has
reconsidered 
its projections of uncertainties on Higgs processes and has
presented and defended
some more optimistic estimates in the 
report  \cite{ATLASnew}.  However, neither
of these  ATLAS papers  includes
estimates for the capability to measure $h\to b\bar b$, which, as
noted
above, plays a central  role in any global fit.  It is not possible to
estimate
the uncertainty in individual Higgs boson couplings without that
information.

CMS presented a less conservative set of estimates, using two
scenarios.
In Scenario 1, current systematic and theoretical errors were used.
In Scenario 2, theoretical errors were halved and  systematic errors
were assumed to 
decrease as the square root of the integrated luminosity.  The
estimates
 under Scenario 2 were meant to express the most optimistic estimates
for the measurement of Higgs couplings, especially at high luminosity.
One might understand them as estimating the {\it opportunity} that 
high-luminosity running of the LHC will make available.   However, 
CMS did not provide the input measurement errors but, rather, only
the results of some global fits.  It is not possible to use this
information 
to carry out other potentially interesting fits or to carry out joint
fits with expected results from other facilities.

For this reason, it is interesting to propose an interpretation of the 
CMS results in terms of a set of errors on a minimal set of
measurements,
assumed to be independent, that reproduces their fit  results.
In Table~\ref{tab:CMSinterp}, I give my proposed interpretation of
the results, along with the ATLAS estimates, from \cite{ATLASnew},
 for comparison.
The estimates given are coherent, in the sense that the accuracy
improves systematically from Scenario 1 at 300~fb$^{-1}$ to 
Scenario 2 at 3000~fb$^{-1}$, but never with a large step.  By the
definition of Scenario 2, the theoretical errors for Scenario 2 at 
3000~fb$^{-1}$ are identical to those at 300~fb$^{-1}$ and the 
experimental errors decrease by $\sqrt{10}$.  I have taken the
theory errors to be the 
errors on the total production cross section as given by the 
LHC Higgs Cross Section Working Group~\cite{Yellow,YellowWeb,Serge}.
It is not possible to obtain the fit results presented by CMS with theory 
errors as large as those quoted by ATLAS in \cite{ATLASnew}.

\begin{table}
\begin{center}
\begin{tabular}{lccc} 
{\bf 300~fb$^{-1}$ : }\\ 
 Observable   & ATLAS & CMS-1 & CMS-2  \\ 
\hline 
$\sigma(gg)\cdot BR(\gamma\gamma)$ &  12 $\oplus$ 19 & 6 $\oplus$ 12.3
& 3 $\oplus$  6.2 \\
$\sigma(WW)\cdot BR(\gamma\gamma)$ &  47 $\oplus$ 15 & 20 $\oplus$ 2.4&
14 $\oplus$ 1.2  \\
$\sigma(gg)\cdot BR(WW)$                       &   8
       $\oplus$ 18  & 6 $\oplus$ 12.3 & 5 $\oplus$ 6.2\\
$\sigma(WW)\cdot BR(WW)$                      &    20 $\oplus$ 8
& 35 $\oplus$ 2.4 & 28 $\oplus$ 1.2\\
$\sigma(gg)\cdot BR(ZZ)$     &  6 $\oplus$ 11        &
   7 $\oplus$ 12.3 
 & 5 $\oplus$ 6.2\\ 
$\sigma(WW)\cdot BR(ZZ)$        & 31 $\oplus$ 13  &  12 $\oplus$     2.4       &
  10 $\oplus$ 1.2\\
$\sigma(gg)\cdot BR(\tau\tau)$  & ---  &  13   $\oplus$  12.3    &
6 $\oplus$ 6.2 \\
$\sigma(WW)\cdot BR(\tau\tau)$               &16 $\oplus$  15
&   16 $\oplus$ 2.4
& 9 $\oplus$ 1.2 \\
$\sigma(Wh)\cdot BR(b\bar b)$     &  ---    &  17 $\oplus$    3.8    &
14 $\oplus$ 1.7  \\
$\sigma(t\bar t h)\cdot BR(b\bar b)$ & ---   & 60 $\oplus$  11.7   & 50  $\oplus$ 5.9 \\
$\sigma(t\bar t h)\cdot BR(\gamma\gamma)$ & 54 $\oplus$  10 & 40
$\oplus$ 11.7 & 38 $\oplus$ 5.9 \\
$\sigma(Zh)\cdot BR(invis) $    &  ---    &  16  $\oplus$    4.3     &
11 $\oplus$ 2.2  \\\hline\hline \\
 & \\ 
{\bf 3000~fb$^{-1}$ :} \\ 
Observable   &   
ATLAS-HL & CMS-HL-1 & CMS-HL-2  \\ 
\hline
$\sigma(gg)\cdot BR(\gamma\gamma)$ &  5 $\oplus$ 19 & 4 $\oplus$ 12.3
&0.9 $\oplus$ 6.2 \\ 
$\sigma(WW)\cdot BR(\gamma\gamma)$ &  15 $\oplus$ 15 & 10 $\oplus$ 2.4 &
4.4 $\oplus$1.2 \\
$\sigma(gg)\cdot BR(WW)$                       &   5
       $\oplus$ 18 &6  $\oplus$12.3  &1.6 $\oplus$ 6.2\\
$\sigma(WW)\cdot BR(WW)$                      & 9 $\oplus$ 8 &
24 $\oplus$ 2.4  & 8.9 $\oplus $ 1.2\\ 
$\sigma(gg)\cdot BR(ZZ)$                         &  4 $\oplus$ 11  &4  $\oplus $ 12.3 &1.6  $\oplus $ 6.2 \\
$\sigma(WW)\cdot BR(ZZ)$         &  16 $\oplus$ 13    & 7 $\oplus $   12.3  &  1.9  $\oplus $ 6.2\\
$\sigma(WW)\cdot BR(\tau\tau)$               & 12 $\oplus$ 15
& 8 $\oplus $ 2.4 & 2.8 $\oplus $ 1.2 \\
$\sigma(Wh)\cdot BR(b\bar b)$    &---   & 8 $\oplus $       3.8     &
4.4 $\oplus $  1.7  \\
$\sigma(t\bar t h)\cdot BR(b\bar b)$  &  ---  & 35 $\oplus $  11.7   & 16
$\oplus $  5.9 \\
$\sigma(t\bar t h)\cdot BR(\gamma\gamma)$ & 17 $\oplus$ 12  &  28 $\oplus$
11.7  & 12 $\oplus $ 5.9 \\
$\sigma(Zh)\cdot BR(invis) $      & ---     &  10 $\oplus $    4.3
&  3.5  $\oplus $   2.2  \\ 
\end{tabular}
\caption{Error estimates for measurement of Higgs boson processes at
  the
LHC. All numbers are given as 1 $\sigma$ uncertainties, in \%.
 Errors are given in the form (experiment)$\oplus$(theory), where
(theory) is an error on the theory used to extract the rate. These
errors are added in quadrature in the analysis. The first three columns
give estimates for 14~TeV with 300~fb$^{-1}$; the second three columns
gives estimates for 14~TeV and 3000~fb$^{-1}$.  The columns for ATLAS
give numbers presented in \cite{ATLASnew}.  The
columns for CMS are my own estimates, justified only by the results of
the
fits shown in Table~\ref{tab:CMSfit}.  CMS-1 denotes Scenario 1; CMS-2
denotes Scenario 2. }
\label{tab:CMSinterp}
\end{center}
\end{table}

In Table~\ref{tab:CMSfit}, I compare the results obtained by applying 
my naive fitting method to the numbers in Table~\ref{tab:CMSinterp}
for the fits discussed in the CMS paper~\cite{CMSWP} with the results
presented in that paper. Three types of data are presented.  The first
is
an error for the determination of the overall rate $\mu$ into the
various
final states.  I obtain these by combining the results for reactions
with 
the same final state in Table~\ref{tab:CMSinterp}.  The second is 
the result of a 7-parameter fit taking $\kappa_c = \kappa_t$ and
$BR_{inv} =BR_{unr}  = 0$.  Finally, I obtain a  limit on the
invisible
branching ratio from an 8-parameter fit with
$BR_{inv}$
taken nonzero and with the restriction 
\beq \kappa_W, \kappa_Z < 1.
\eeq{WZrestrict}
  This 
latter constraint~\cite{GHW} is a minimally model-dependent
 approach to fixing
the Higgs width that is often used in fits to Higgs couplings at 
hadron colliders~\cite{myHiggsBR,Duhrssen,Lafaye,Klute}.   This 
follows the methodology used by CMS, as explained in the 
Snowmass Higgs working group report~\cite{Higgsworking}. Following 
the prescription given there, the quoted
uncertainties
on $BR_{inv}$ do not include the direct constraint from measurement 
of $pp\to Zh$.   This weak constraint, however, has only a small 
effect on the fit.

The uncertainties in the total cross
sections
reported in ~\cite{Yellow,YellowWeb} are large for $gg$ fusion and
associated production of the Higgs with $t\bar t$ but quite small 
for vector boson fusion and for associated production with $W$ and
$Z$.   This can be seen in the theory errors reported for CMS in 
Table~\ref{tab:CMSinterp}.
As the luminosity increases, it is possible to decrease the theory
error 
in output Higgs couplings by 
relying increasingly on measurements of Higgs production in these 
latter two modes.  
The evolution of the $\mu$ and $\kappa$ accuracies reported by 
CMS from 300~fb$^{-1}$
to 3000~fb$^{-1}$ reflects increasing reliance at higher luminosity on
the vector boson fusion production mode.  
It is very important to note this special role
of vector boson fusion in any considerations of the experimental
program of the High-Luminosity LHC.  The model I have
presented
captures that this evolution to higher accuracies, at least in a
qualitative way.

\begin{table}
\begin{center}
\begin{tabular}{l|cc|cc} 
$\mu$ values &  300~fb$^{-1}$  & & 3000~fb$^{-1}$ &  \\ 
  & CMS & here  & CMS &  here  \\ \hline
 $\gamma\gamma$  &  [ 6 , 12 ] & [ 6.2 , 11.3 ]   &   [ 4 , 8 ]  & [
 3.7 , 8.0 ] \\
 $WW$ &  [ 6 , 11 ] & [ 7.6 , 12.7]  &  [ 4 , 7 ] & [ 5.2 , 11.9 ] \\
$ZZ$ &   [  7 , 11 ] &  [ 6.2 , 12.7 ]  &  [ 4 , 7 ] &  [ 3.0 , 7.0 ] \\
$b\bar b$ & [ 11 , 14 ]  & [ 13.6 , 16.7 ]  & [ 5 , 7 ] & [ 4.7 , 8.6 ] \\
$\tau^+\tau^-$ &  [ 8 , 14 ] & [ 6.2 , 12.0 ]  &  [ 5, 8 ] &  [ 2.8 , 7.2 ]  \\
invis. &  [ 11 , 17 ]  &  [11.2 , 16.6 ]    &   [ 4 , 11 ]  & [ 4.1 ,
10.9 ] \\
\hline
$\kappa$ values &  300~fb$^{-1}$  & & 3000~fb$^{-1}$ &  \\
  & CMS & here   & CMS &  here  \\ \hline
$\gamma$ &   [ 5 , 7 ]  &[ 5.7 , 9.0 ]   & [ 2 , 5 ] &  [ 2.9 , 6.5 ]   \\
$W$&  [ 4 , 6 ] & [ 4.2 , 5.4 ] & [  2 , 5 ]  & [ 1.6 , 3.3 ] \\ 
$Z$&   [ 4 , 6 ] & [ 5.7 , 8.5 ]  &  [ 2 , 4 ] & [ 2.8 , 6.3 ]  \\
$g$&   [ 6 , 8 ]  &  [ 4.9 , 6.9 ] &  [ 3 ,  5 ] & [ 2.3 , 4.8 ] \\
$b$&    [ 10 , 13 ] & [ 11.4 , 14.9 ] & [4 , 7 ] & [ 4.2 , 8.5 ] \\
$t$&      [ 14 , 15 ] & [ 17.3 , 20.5 ] & [ 6 , 8 ] & [ 5.7 , 12.9 ]  \\
$\tau$&   [ 6 , 8 ] & [ 5.8 , 9.5 ]  &  [ 2 , 5 ] & [ 2.7 , 6.5 ] \\
inv.&    [ 8 , 11 ]  & [ 6.3 , 8.0 ] & [ 4 , 7 ] & [ 2.0 , 4.0 ]  \\
\end{tabular}
\caption{Comparison of the results of fits with the inputs in
  Table~\ref{tab:CMSinterp} to the fit results given in
  \cite{CMSWP}. All numbers are given as 1 $\sigma$ uncertainties, in \%.  In expressions in
brackets,
the first entry is for Scenario 2, the second is for Scenario 1.}
\label{tab:CMSfit}
\end{center}
\end{table}
%

There are some defects in the agreement of my model with the CMS
results.  The most serious is the constraint on the invisible modes of 
Higgs decay, which is significantly stronger in my fit than that
reported
in \cite{CMSWP}.   This may be the result of my treating correlated
theoretical errors as  uncorrelated, which stiffens the global  pattern
of the constraints.  The 
CMS analysis also uses a much larger number of input measurements,
with correspondingly larger errors, and 
takes proper account of the correlations among these errors.
Such a treatment is beyond the level of my 
interpretation.  Nevertheless, I hope that the information
that I have given in Table~\ref{tab:CMSinterp}
 will suffice for the purpose of estimating Higgs 
capabilities for experiments that will be carried out in the future.

\section{Fits including invisible modes}

The coupling accuracies listed in Table~\ref{tab:CMSfit} come mainly
from a fit in which only SM modes of Higgs decay are taken into
account.
Only the uncertainties quoted for the invisible branching fraction
are based on a fit that allows
the Higgs to decay invisibly.   It is interesting to perform
another
fit to understand how the possible presence of invisible modes of 
Higgs decay affects the capabilities of the LHC experiments.

As noted above, there are two types of  ``invisible'' modes of Higgs decay
at  the LHC.   In Section 2, I distinguished between {\it invisible}
modes
of Higgs decay (which can actually be measured directly at the LHC)
from {\it unrecognizable} modes of Higgs decay, which are hidden 
by Standard Model backgrounds.

To take account of both possibilities, we can fit the inputs given in 
Table~\ref{tab:CMSinterp} with a model in which $\kappa_c  = 0$ but
 all of the other 9 parameters in
\leqn{totkappa}
are allowed to float.  This fit includes the constraint on $Zh$ 
production with the Higgs decaying invisibly.
 The decay of the Higgs to $c\bar c$ is an
unrecognizable
mode in the sense of the previous section, so I omit this variable,
as explained in the definitions of the parameters of  \leqn{kappahdef}.
 To
keep the total width of the Higgs boson from running to large values,
we must impose the condition $\kappa_W, \kappa_Z < 1$.   The
prescription is close to the one 
proposed in \cite{myHiggsBR}  to measure the ability of colliders to 
perform model-independent determinations of the Higgs couplings.

The results of the fit are shown in Table~\ref{tab:nineparam}. The
table
compares these results to those of the 7-parameter fit including
only Standard Model decays  discussed in the previous section.

\begin{table}
\begin{center}
\begin{tabular}{l|cc|cc} 
300~fb$^{-1}$  & Scenario 2  & & Scenario 1 &  \\
  & 7-param  & 9-param & 7-param &  9-param  \\ \hline
$\gamma$ &  5.7 &   4.3     &   9.0   & 7.3 \\
$W$& 4.2  & 3.5 &  5.4 &  4.6     \\ 
$Z$&  5.7  &  5.0  & 8.5 &   6.6  \\
$g$& 4.9  & 4.1   & 6.9  &    6.3\\
$b$&  11.4 & 7.6  & 14.9 &  10.2 \\
$t$&   17.3  & 17.3 &  20.5  &  20.6   \\
$\tau$&  5.8 &4.4 &  9.5 &  7.7  \\
invis.&   ---&  4.6 &  --- &   6.1   \\ \hline 
3000~fb$^{-1}$  & Scenario 2  & & Scenario 1 &  \\
  & 7-param  & 9-param & 7-param &  9-param  \\ \hline
$\gamma$ &  2.9  &2.4 &  6.5  & 5.3   \\
$W$& 1.6  & 1.2  &   3.3  &  2.3 \\ 
$Z$&  2.8   &   2.2  &   6.3 & 4.3 \\
$g$&   2.3   & 2.0  &    4.8  & 4.4 \\
$b$&   4.2  &  2.9 &   8.5 & 6.0 \\
$t$&    5.7 & 5.6  &  12.9 & 12.8  \\
$\tau$&  2.7  &2.2  &   6.5 & 5.1  \\
invis.&   --- & 1.5    &   ---& 3.2  \\
\end{tabular}
\caption{Comparison of the results for Higgs coupling 
uncertainties, in \%, 
without and with allowance for invisible and unrecognized Higgs
decays, as defined in the text.  The analyses are 
performed, respectively, with 7-parameter fits including only Standard
model decay modes and constrained 9-parameter fits including invisible and 
unrecognizable modes,
 as described in the
text.  All numbers are computed with the inputs in
  Table~\ref{tab:CMSinterp} and are given as 1 $\sigma$ uncertainties,
in \%.
The lines labeled $invis$ refer to invisible modes of Higgs
boson decay.}
\label{tab:nineparam}
\end{center}
\end{table}
%

It is noteworthy that the coupling errors in the 9-parameter fit are
typically
smaller than those in the 7-parameter fit, despite the fact that the 
9-parameter fit contains extra unconstrained degrees of freedom.  This is
the effect of imposing the condition $\kappa_W, \kappa_Z < 1$.   A
 fit with this condition imposed but with no allowance for
invisible or
unrecognized modes gives similar results for the uncertainties in the 
$\kappa_A$ for Standard Model decay  modes. 

\section{Estimates of coupling accuracy for ILC}

Using a similar methodology, it is straightforward to produce
estimates of the accuracy of the Higgs boson couplings that will be
obtained at the various stages of the ILC.  The ILC Higgs White
Paper~\cite{ILCHiggs}
reviews and improves the Higgs boson coupling analyses reported
in the ILC TDR~\cite{ILCTDR}.  This paper also  emphasizes that the 
long-term program of the ILC will lead to further improvements in 
Higgs coupling measurements.  Whereas the LHC Higgs coupling
determinations become dominated by systematic errors  in the
High-Luminosity LHC era, the ILC measurements are always 
dominated by statistical errors  and so improve continually
with larger data samples.  Running at any $\ee$ center of mass 
energy makes a contribution, especially because the cross 
section for $WW$ fusion production of Higgs grows with 
energy.  The program set out in \cite{ILCHiggs}
includes the programs of ILC running at 250 and 500~GeV discussed 
in the ILC TDR, the anticipated energy upgrade to 1000~GeV, and also 
a set of luminosity upgrades for which the strategies are mapped out
 in the TDR.  With the 
somewhat conservative luminosity projections of the TDR, this 
program would require 18 Snowmass years ($18\times 10^7$ sec),
comparable to the expected running period of the LHC.  With 
insights gained from the 
experience of operating the ILC, this time could be shortened or 
more integrated luminosity could be obtained. 

In this paper, I will consider the ILC program as progressing 
along the line given in Table~\ref{tab:ILCprogram}.  At each 
successive stage, the new measurements obtained from 
the data sets shown
in the Table are added to all previous ILC measurements. I will 
consider the ILC stages as being carried out in the order given
in Table~\ref{tab:ILCprogram}.   This is a slightly different order
from that assumed in \cite{ILCHiggs}.  It  allows
the full program at 250~GeV and 500~GeV to be completed
in parallel with the construction needed for ILC collisions
at 1000~GeV.   The program through the 500up stage 
would require 12 Snowmass years of data-taking, again 
assuming that  our current understanding of the operation
of a linear collider does not improve after years of running
the ILC.

 Data from the LHC are not included in the fits described in this
section.  Joint fits to LHC and ILC data 
are considered in the next section.

\begin{table}
\begin{center}
\begin{tabular}{l|c|c|c|c|c|c} 
                 & 250 &  500  & 250up  &  500up  &  1000 & 1000up \\ \hline
Energy (GeV)    &    250   &  500 &  250 & 500 &  1000 & 1000 \\ \hline
Luminosity (fb$^{-1}$)& 250 &   500 &   1150 &  1600 &  1000 &  2500
\\ 
\end{tabular}
\end{center}
\caption{The ILC program envisioned in~\cite{ILCHiggs}.
The stages are carried out sequentially, each one adding the 
data set given in the column.  }
\label{tab:ILCprogram}
\end{table}

The fits to ILC data includes measurement of the total 
cross section  $\sigma(\ee\to Zh)$, and independent
measurements of    $BR(h\to b\bar b)$ and 
$\sigma(\ee\to \nu\bar\nu h)BR(h\to b\bar b)$, from 
which the total cross section for the $WW$ fusion process
can be extracted.   The first two of these measurements
are enabled by observing the decaying $Z$ boson  that tags Higgs
boson production in $\ee\to Zh$.   These reactions allow 
the total width of the Higgs boson to be constrained in a 
model-independent way.   Then we can include the possibility
of invisible Higgs decays into the fit without any need for  the 
restriction \leqn{WZrestrict}.

\begin{table}
\begin{center}
\begin{tabular}{l|ccccc} 
         & 250 &  500  &   500up  &  1000 & 1000up \\ \hline
$W$&               4.6   &      0.46       &      0.22        &      0.19   
  &        0.15                                  \\ 
$Z$&               0.78    &      0.50        &     0.23       &       0.22
  &        0.22                                  \\ 
$g$&                6.1    &     2.0        &     0.96        &    0.79
  &         0.60                                 \\ 
$\gamma$ &   18.8 &      8.6       &      4.0       &   2.9
  &           1.9                               \\ 
$b$&                 4.7   &     0.97        &     0.46        &    0.39
  &            0.32                              \\ \
$c$ &               6.4   &      2.6        &    1.2       &    0.98
  &          0.72                                \\ 
$\tau$&          5.2    &     2.0       &       0.89      &   0.79
  &         0.65                                \\ 
invis.&          0.54   &      0.52        &      0.22        &    0.22
  &          0.21                                \\ \hline 
\end{tabular}
\caption{Comparision of the results for Higgs
coupling uncertainties,  in \%,
from ILC at its various stages, based 
on estimates of $\sigma$ and $\sigma \times BR$ 
accuracy given in \cite{ILCHiggs}.   The 
results are based on a 10-parameter fit defined in the 
text.  The stages of the ILC program are those defined
in Table~\ref{tab:ILCprogram}.}
\label{tab:ILCstages}
\end{center}
\end{table}

It should be noted that the fit used here
assumes more information than is included in the corresponding
multi-parameter fits performed in \cite{Higgsworking}, where the 
total width of the Higgs is taken to be an additional free parameter.
I assume that it is possible use the tagged Higgs decays
from $\ee\to Zh$  to put an experimental bound
on exotic Higgs decay modes equal to the direct experimental 
constraint that will be placed on invisible decay modes.  Since an
exotic decay that is not invisible has an observable component, this 
is a 
quite conservative assumption.   I will discuss this assumption 
further in Section 7. 

My estimates will then be based on a 10-parameter fit with
all 10 parameters in \leqn{totkappa} free to vary under the 
constraints given by the ILC measurements.  The underlying
values are assumed to be those of the Standard Model.
The constraint \leqn{WZrestrict} is {\it not} applied.
The measurement
accuracies assumed are those given in Tables~5.4 and 5.5 of
\cite{ILCHiggs}, with exotic modes assumed to have, independently,
 the same 
upper limit  as the  invisible modes. 

\begin{figure}[p]
\begin{center}
\includegraphics[width=0.48\hsize]{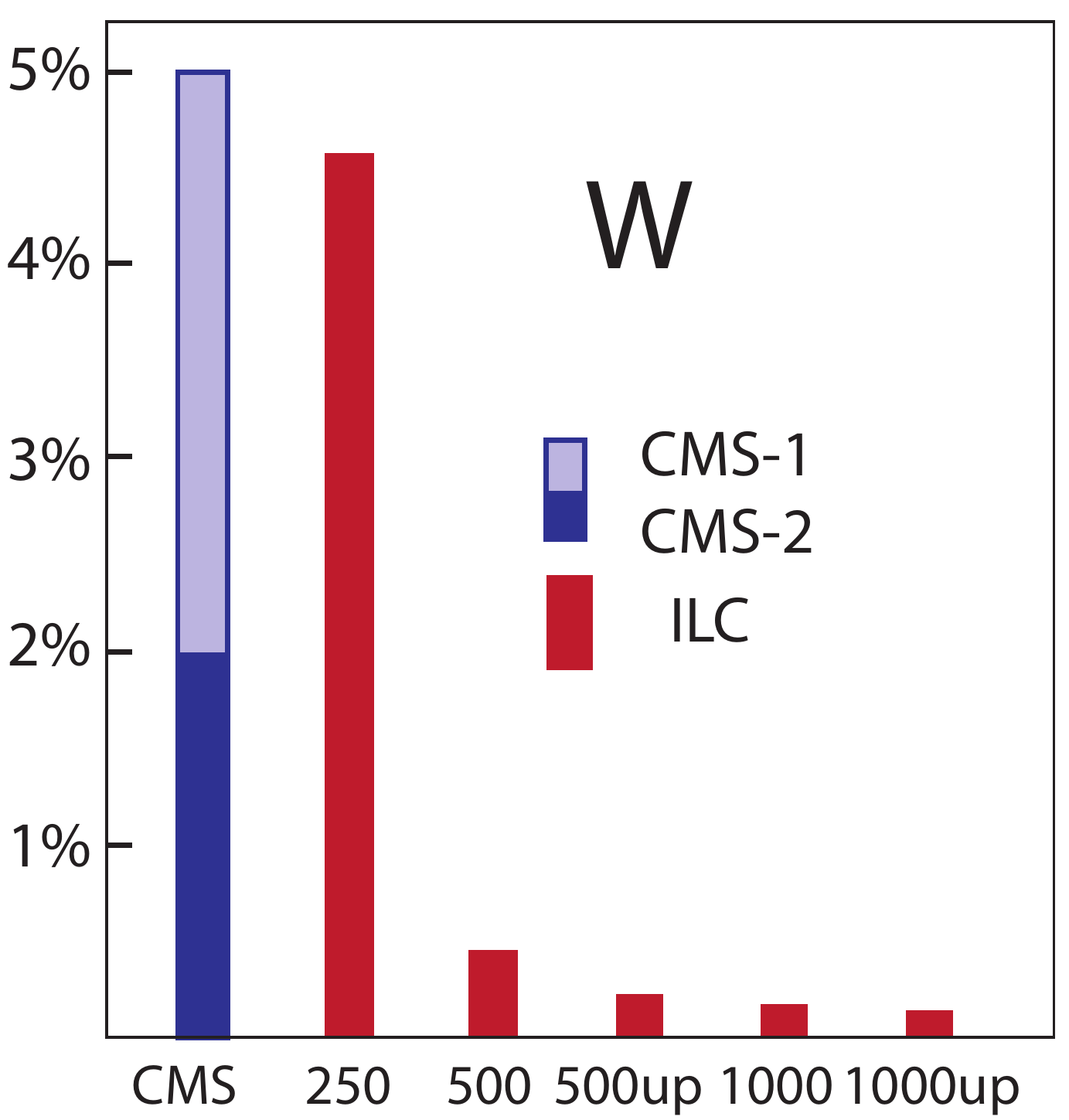}\ \ 
\includegraphics[width=0.48\hsize]{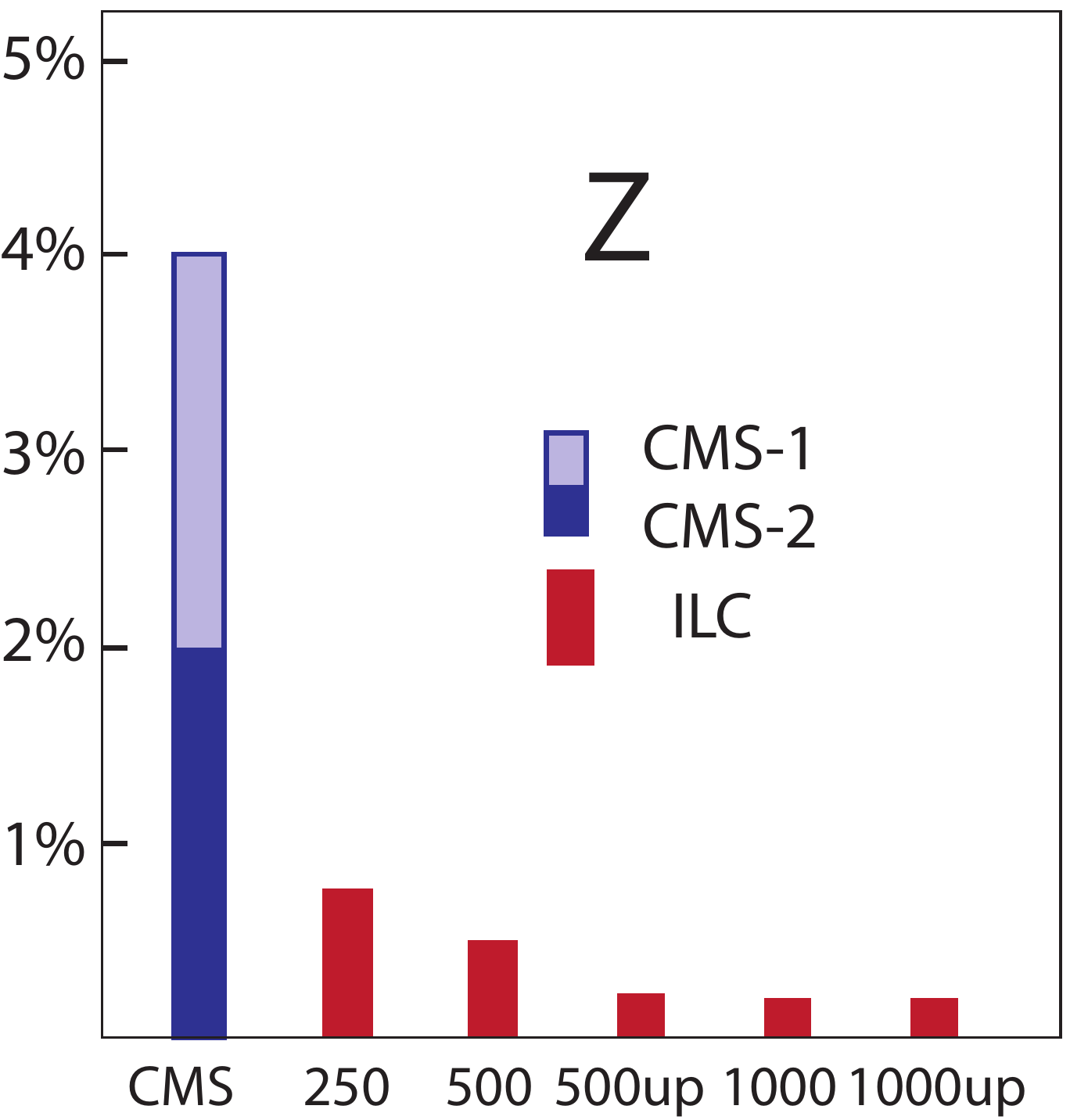}
\end{center}
\caption{Estimates of the ILC measurement accuracies for
the Higgs boson couplings to $WW$ and $ZZ$.  These 
estimates are based on the 10-parameter fit described in 
the text. The 
successive entries correspond to the stages of the ILC
program shown in Table~\ref{tab:ILCprogram}.  The CMS
Scenario 1 and Scenario 2 estimates for 3000~fb$^{-1}$, from 
\cite{CMSWP}, are shown on the 
left.}
\label{fig:WZ}

\bigskip

\bigskip

\begin{center}
\includegraphics[width=0.48\hsize]{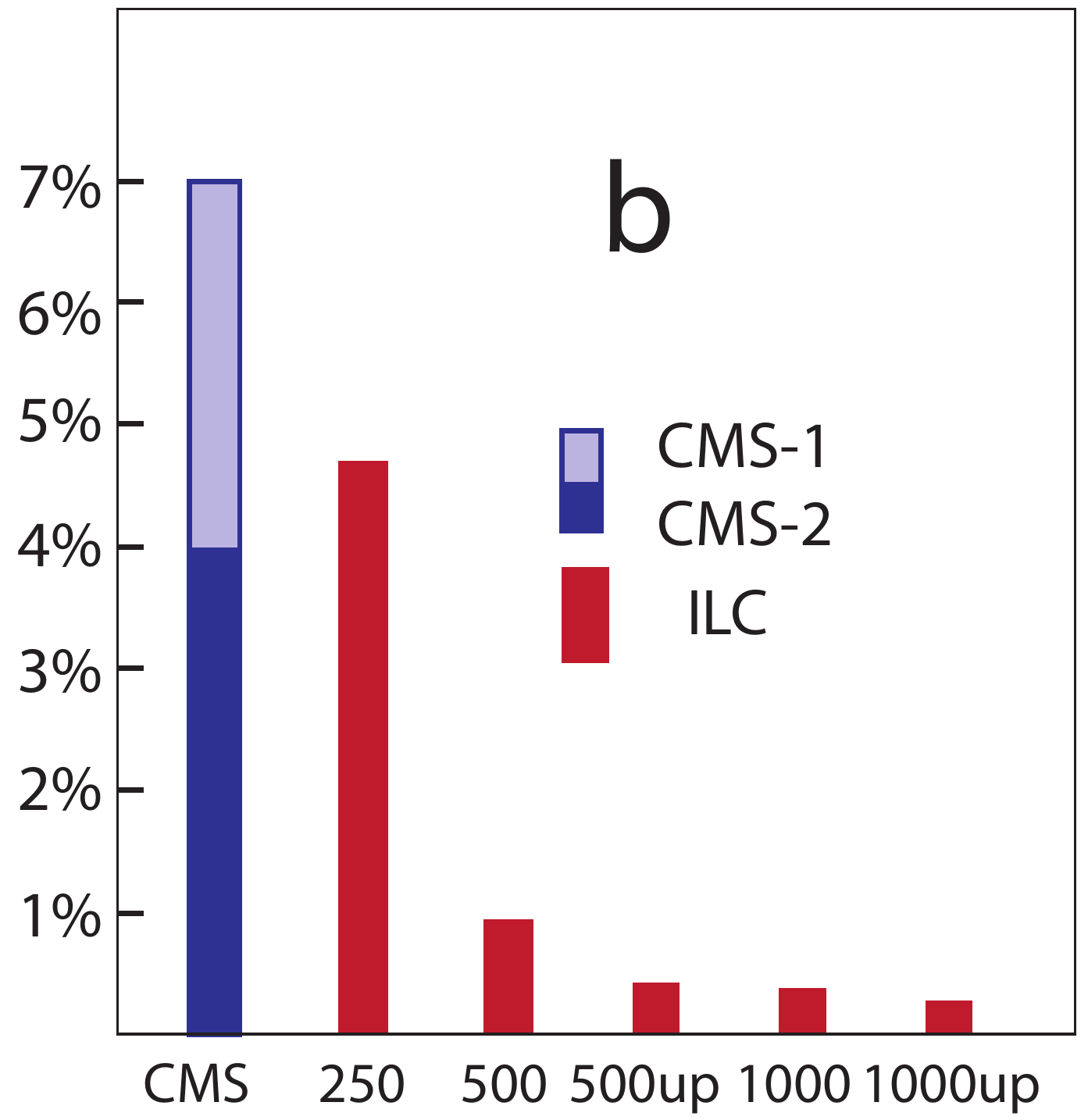}\ \ 
\includegraphics[width=0.48\hsize]{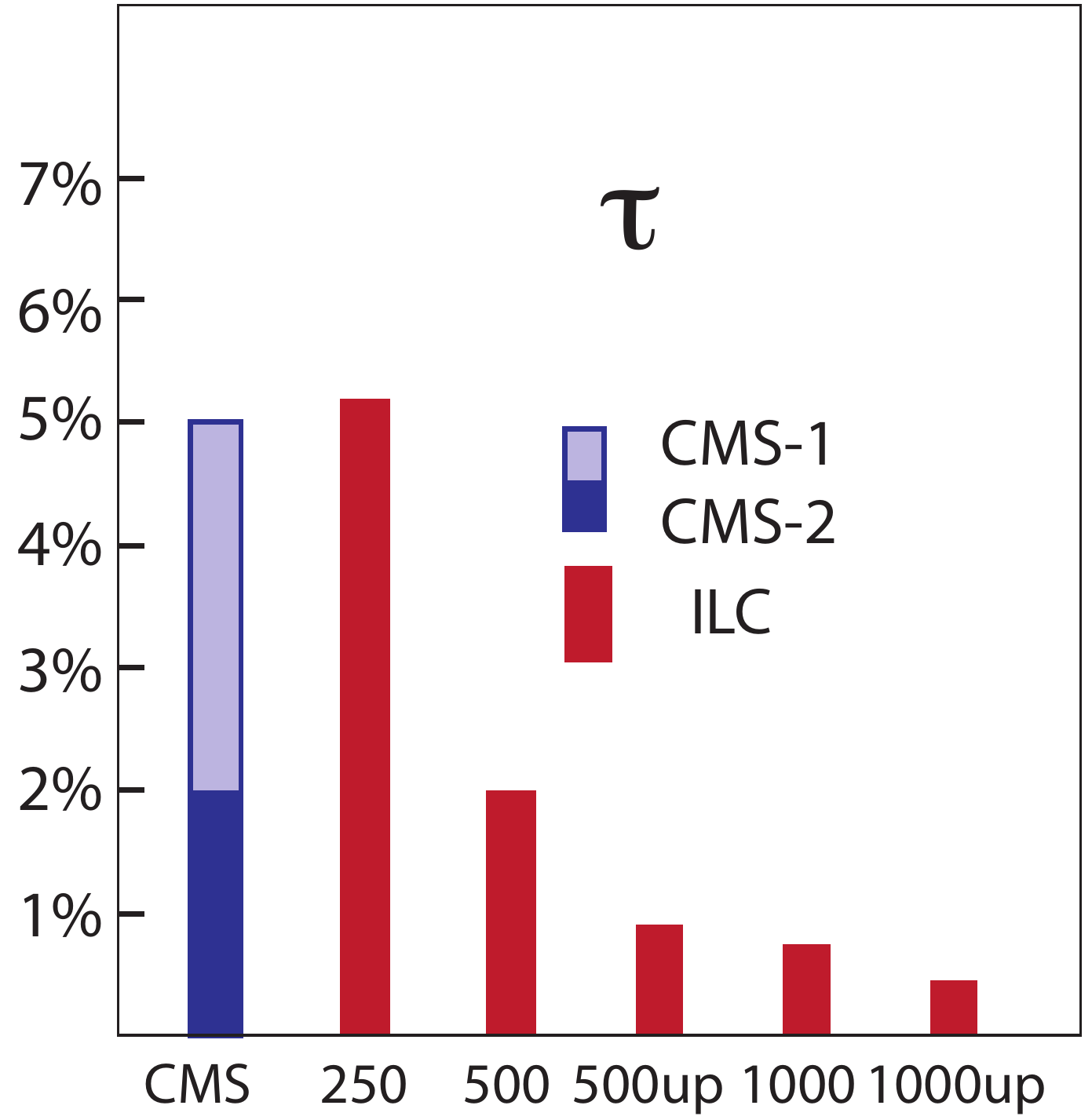}
\end{center}
\caption{Estimates of the ILC measurement accuracies for
the Higgs boson couplings to $b\bar b$ and $\tau^+\tau^-$.  These 
estimates are based on the 10-parameter fit described in 
the text. The 
successive entries correspond to the stages of the ILC
program shown in Table~\ref{tab:ILCprogram}.   The CMS
Scenario 1 and Scenario 2 estimates for 3000~fb$^{-1}$, from 
\cite{CMSWP}, are shown on the 
left.}
\label{fig:btau}
\end{figure}

\begin{figure}[t]
\begin{center}
\includegraphics[width=0.48\hsize]{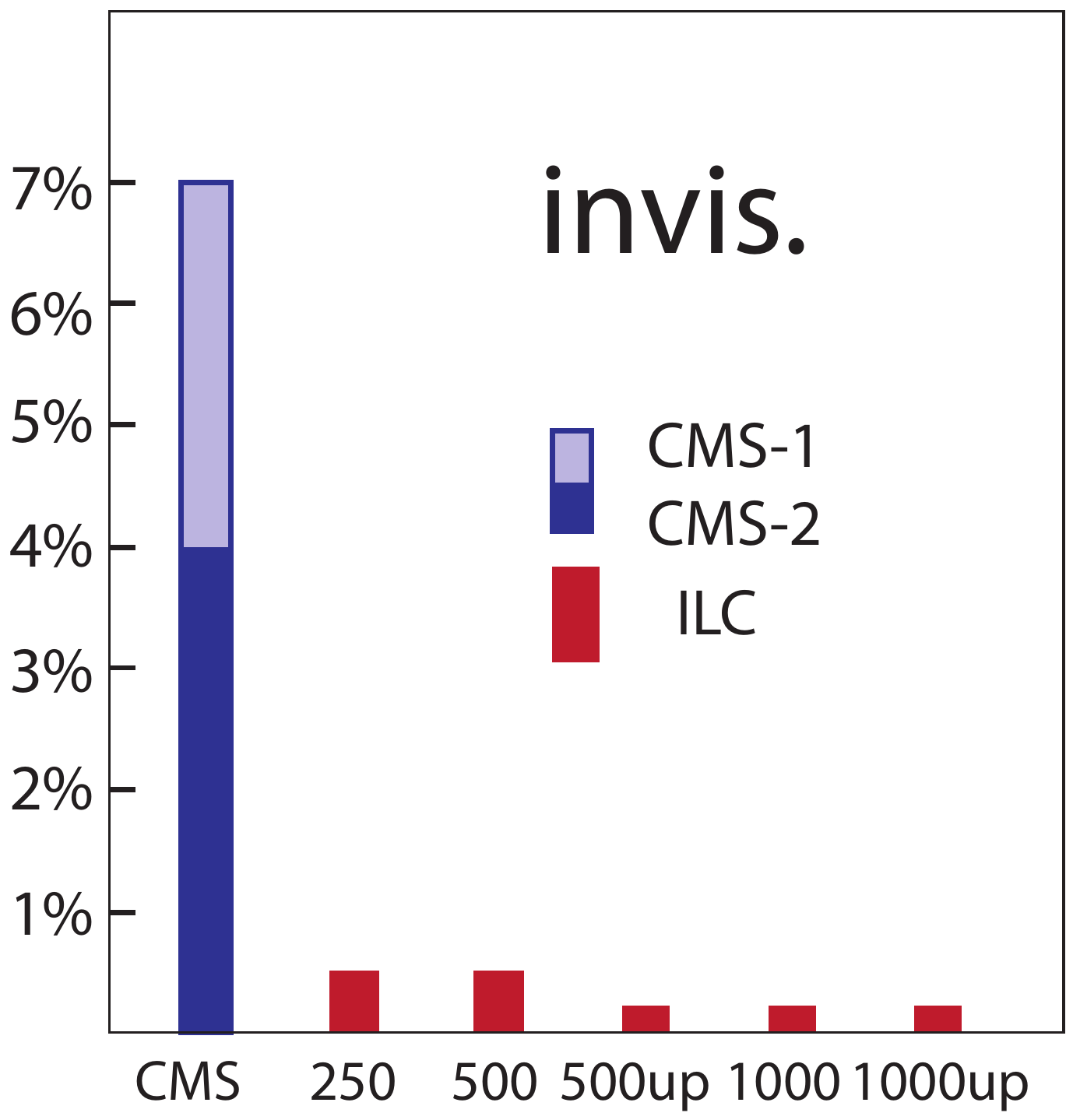}\ \ 
\includegraphics[width=0.48\hsize]{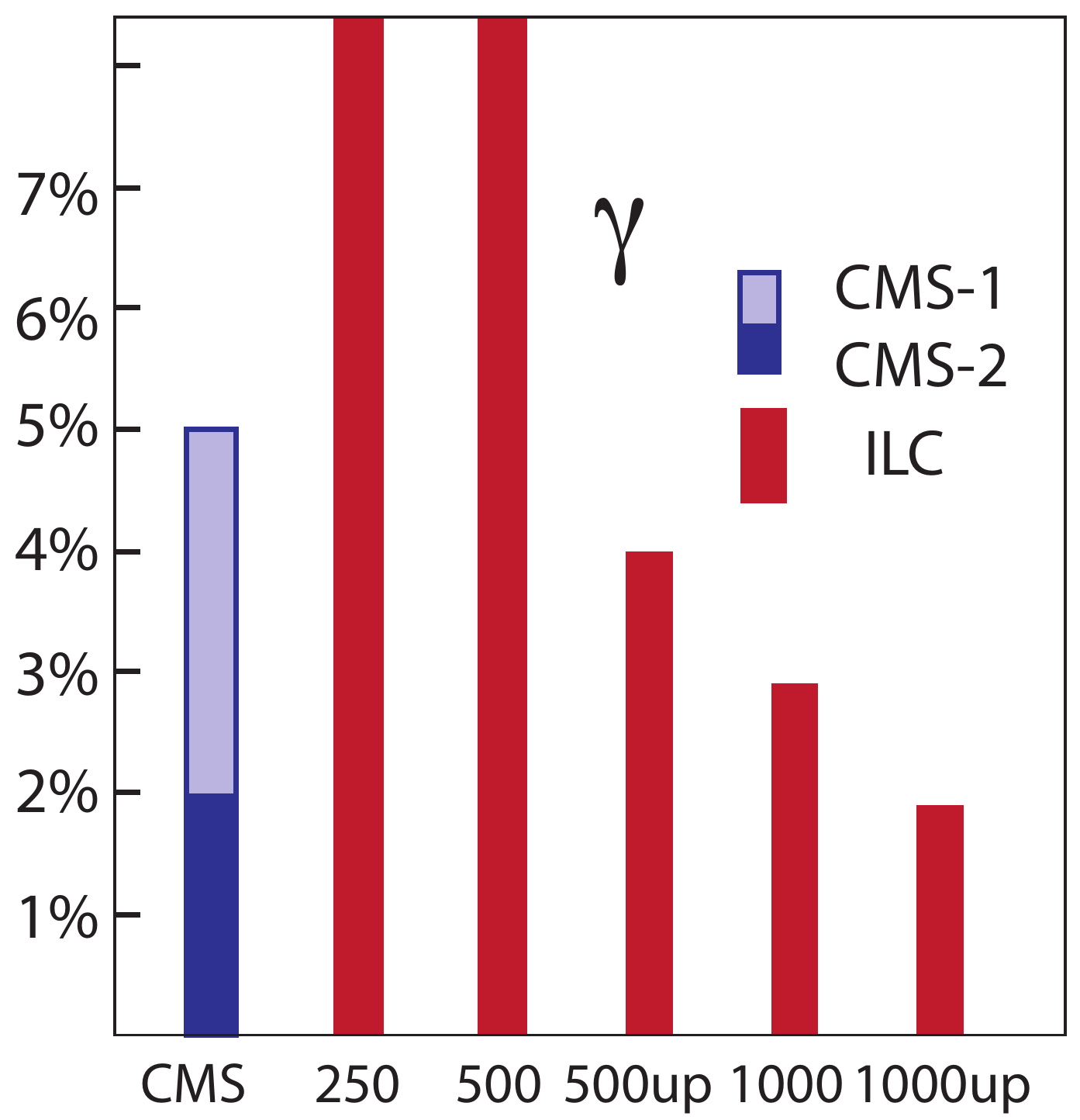}
\end{center}
\caption{Estimates of the ILC measurement accuracies for
the Higgs boson couplings to invisible modes and to 
$\gamma\gamma$.  These 
estimates are based on the 10-parameter fit described in 
the text. The 
successive entries correspond to the stages of the ILC
program shown in Table~\ref{tab:ILCprogram}.  The CMS
Scenario 1 and Scenario 2 estimates for 3000~fb$^{-1}$, from 
\cite{CMSWP}, are shown on the 
left.}
\label{fig:invgam}
\end{figure}

My results for the uncertainties in ILC Higgs coupling determination
at the various ILC stages are given in Table~\ref{tab:ILCstages}. 
These uncertainties are smaller than those estimated in the reports
\cite{Higgsworking} and \cite{ILCHiggs} because I take into account
that the experiments will search for exotic Higgs decays and, if these
are not present, will present strong upper limits.
Graphical comparison of the uncertainties from my analysis
 with those estimated by 
CMS in \cite{CMSWP} for 3000~fb$^{-1}$  are shown for the $WW$ and $ZZ$ couplings
 in Fig.~\ref{fig:WZ}, for the $b\bar b$ and $\tau^+\tau^-$ 
couplings in Fig.~\ref{fig:btau}, and the invisible and $\gamma\gamma$
couplings in Fig.~\ref{fig:invgam}.   All of these estimates 
except for the $\gamma\gamma$ case show a steady progression
to smaller errors with increasing statistics that quickly reach
projections of sub-1\% accuracy.    The conclusion that the 
error on the $\gamma\gamma$ coupling does not achieve high
accuracy will be reconsidered in the next section.

The capabilities of the ILC at 1000 GeV for direct measurements of 
the $ht\bar t$ coupling and the Higgs self-coupling do not enter the
analysis I have presented here.  They are nevertheless impressive. 
 Those measurements are 
described in detail in  \cite{ILCHiggs}.

\section{Joint fits to LHC and ILC data}

It is an interesting exercise to fit ILC data in isolation, but a more 
realistic projection would take into account the fact that the 
LHC results for 3000~fb$^{-1}$ will be known at the time of all
except possibly the earliest stages of the ILC program.   The 
figures shown in the previous section emphasize the competition
between $\ee$ and $pp$ colliders, but there is also the possibility 
of synergy.

It is not possible carry out joint fits for the ILC and CMS data from 
the information provided in \cite{CMSWP}.  However, my interpretation
of the CMS results as a set of uncertainties for $\sigma \cdot BR$ 
measurements makes it possible to estimate the improvement in the 
complete picture of Higgs couplings that will come from combining
the ILC and LHC results.  It is straightforward to perform
 the 10-parameter fit described
in the previous section with the additional input of the projected
CMS results from the column CMS-HL-2 of
 Table~\ref{tab:CMSinterp}.  The resulting uncertainties are shown in 
Table~\ref{tab:ILCplusLHC}.
For each entry in this table, the first column gives the estimated
uncertainty from Table~\ref{tab:ILCstages} and the third column 
gives the estimated uncertainty when the data from the 
ILC program, to this stage,
is combined with the data from the LHC program with 3000~fb$^{-1}$
assuming the CMS Scenario 2 errors.

When we compare  these pairs of numbers, it is apparent that the 
main effect of this combination is a dramatic improvement of the 
uncertainty on the Higgs coupling to $\gamma\gamma$.  This 
impact is clarified if we take a different approach. In \cite{ATLASWP}
and \cite{ATLASnew}, the ATLAS Collaboration presents 
projected uncertainties on ratios of Higgs branching ratios.
In many cases, these ratios of branching ratios have substantial
theoretical and modelling errors.  For example, in the measurement
of $BR(WW^*)/BR(ZZ^*)$, a jet veto is used for the identification 
of $WW^*$ 
events but not for $ZZ^*$ events.  However, there is one ratio that 
should be almost completely free of theoretical errors.  This is 
the ratio  $BR(\gamma\gamma)/BR(ZZ^*)$.   For both the
$\gamma\gamma$ and the $ZZ^*$ 
final states, the dominant contribution to the measurement 
comes from $gg$ production.  But also, more importantly, 
both of these final states allow 
the Higgs boson to be completely reconstructed, so that
it is possible to tailor the measurement in such a way that the 
Higgs production dynamics is identical for the two samples 
being compared.  In \cite{ATLASWP}, ATLAS claims that, with
3000~fb$^{-1}$, this 
measurement could  be made with an uncertainty of 2.9\%, with 
{\it no} theoretical component.   In \cite{ATLASnew}, this 
estimate is revised to an experimental component of 3.6\%
plus a theoretical component of 13.8\% (1.8\% and 7.9\%, respectively,
in the ratio of couplings $\lambda_{Z\gamma}$).  In my opinion, the 
inclusion of this theoretical component is certainly an error~\cite{Feng}.

Motivated by these considerations, I have carried out the 10-parameter
fit to ILC results combined with the single result from 
the High-Luminoisity LHC that the ratio of branching ratios
$BR(\gamma\gamma)/BR(ZZ^*)$ is measured to 3.6\%.   These 
are the results given in the second column for each entry in 
Table~\ref{tab:ILCplusLHC}.   The combined results in this case are
comparable to those obtained from the combination with the
full set of 
CMS projections. 

\begin{figure}[t]
\begin{center}
\includegraphics[width=0.48\hsize]{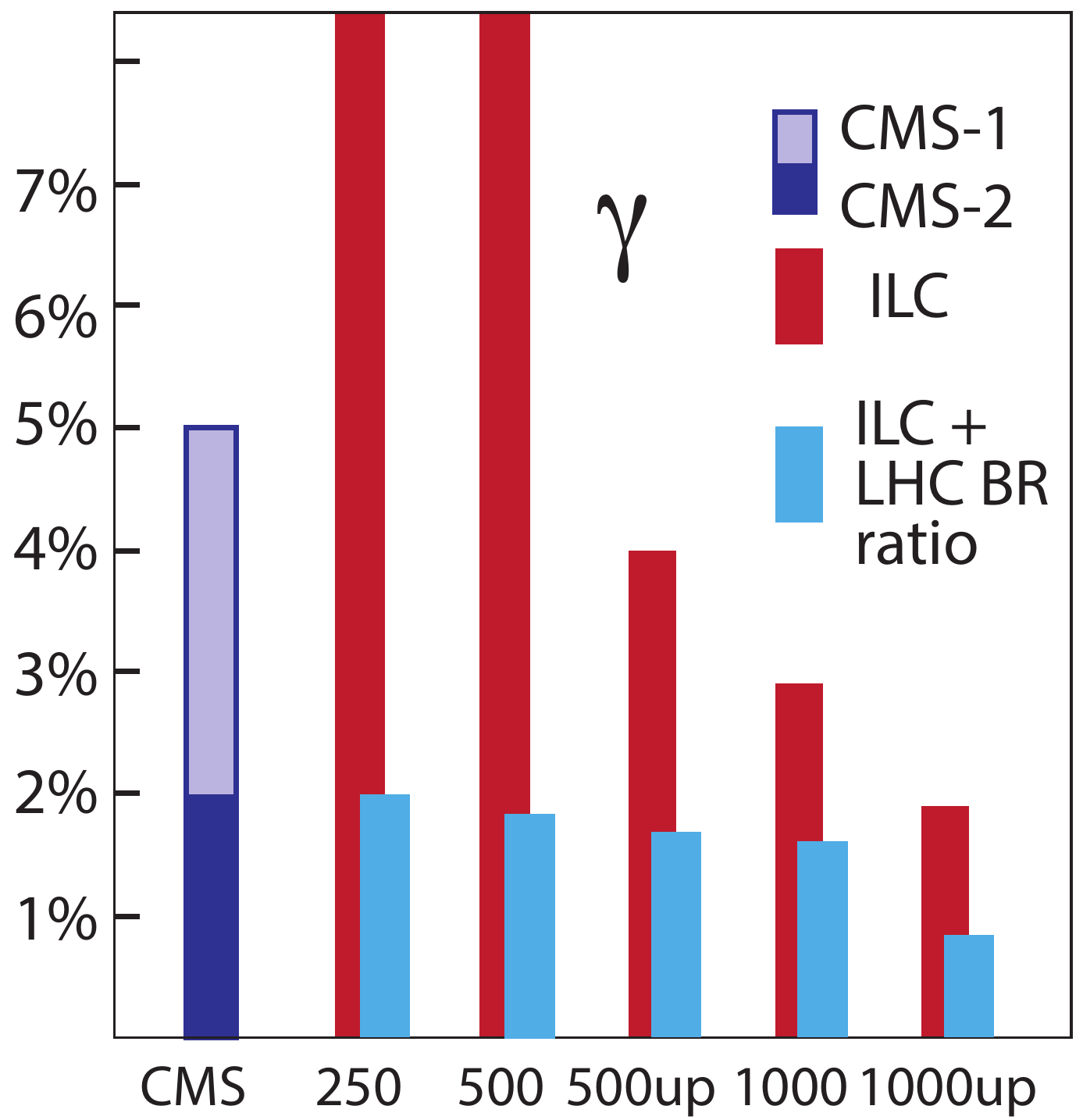}
\end{center}
\caption{Estimates of the ILC measurement accuracies for
the Higgs boson couplings to
$\gamma\gamma$   when combined with the 
measurement of $BR(\gamma\gamma)/BR(ZZ^*)$ 
projected by ATLAS~\cite{ATLASnew}.     The 
successive entries correspond to the stages of the ILC
program shown in Table~\ref{tab:ILCprogram}.   The CMS
Scenario 1 and Scenario 2 estimates for 3000 fb$^{-1}$, from 
\cite{CMSWP}, are shown on the 
left.}
\label{fig:invgammawATLAS}
\end{figure}

The revised estimates for the uncertainties in the Higgs coupling 
to $\gamma\gamma$ from the various ILC stages
are displayed in Fig.~\ref{fig:invgammawATLAS}.   The eventual 
error on the $\gamma\gamma$ coupling is somewhat better than
1.8\% in the 500 GeV ILC era and becomes significantly better,
even below 1\%,  using
the statistics from the $WW$ fusion reaction at the ILC in the
1000~GeV
era.   In comparing the results from CMS and the combined 
ILC/LHC analysis, it is important to remember that the 
former is based on a model-dependent fit while the latter
is model-independent and dominated by statistical errors.

\begin{table}
\begin{center}
\begin{tabular}{l|ccc|l|ccc} 
 250   & ILC &  w BR  &  w CMS-2  &   & ILC  & w BR & w CMS-2 \\ \hline
$W$&4.6   &  4.6 & 1.4  & $Z$ & 0.78  & 0.77  & 0.57  \\ 
$g$&6.1  &   6.0 & 2.0 & $\gamma$ & 18.8  &   2.0 & 2.0  \\ 
$b$& 4.7  & 4.5 & 1.8 & $c$ & 6.4 &6.3  & 4.6  \\
$\tau$& 5.2  & 5.0 &  1.6 & invis.&  0.54 & 0.54   &   0.52 \\  \hline\hline
500   & ILC &  w BR  &  w CMS-2  &   & ILC  & w BR & w CMS-2 \\ \hline
$W$& 0.46  & 0.46  & 0.43  & $Z$ & 0.50  & 0.50  &  0.47 \\ 
$g$& 2.0 &2.0  & 1.4 & $\gamma$ & 8.6  &  1.8 & 1.9 \\ 
$b$&  0.97 &0.96  &  0.80 & $c$ & 2.6 & 2.6  &  2.5 \\
$\tau$& 1.9  & 1.9 & 1.3  & invis.& 0.52  & 0.52 & 0.51  \\  \hline\hline
500up  & ILC &  w BR  &  w CMS-2  &   & ILC  & w BR & w CMS-2 \\ \hline
$W$&  0.22 &  0.22 & 0.21   & $Z$ &  0.23  & 0.23  &  0.23 \\ 
$g$& 0.96 &  0.96 & 0.85 & $\gamma$ & 4.0  &1.7  & 0.9 \\ 
$b$&  0.46 & 0.46  & 0.43 & $c$ & 1.2 &  1.2 & 1.2  \\
$\tau$& 0.89  &  0.88 & 0.78  & invis.& 0.22 & 0.22 & 0.22  \\  \hline\hline
 1000  & ILC &  w BR  &  w CMS-2  &   & ILC  & w BR & w CMS-2 \\ \hline
$W$&0.19   &  0.19  & 0.19  & $Z$ &  0.22 & 0.22  & 0.22  \\ 
$g$& 0.79 & 0.79 & 0.72 & $\gamma$ & 2.9  & 1.6 & 0.89 \\ 
$b$& 0.39  &  0.39 &  0.37 & $c$ &0.98  & 0.97 &  0.98 \\
$\tau$& 0.79  & 0.79 & 0.70  & invis.& 0.22  & 0.21 & 0.21  \\  \hline\hline
1000up  & ILC &  w BR  &  w CMS-2  &   & ILC  & w BR & w CMS-2 \\ \hline
$W$& 0.15  & 0.15  & 0.15  & $Z$ &  0.22  & 0.22  &  0.21 \\ 
$g$& 0.60  & 0.56 & 0.56  & $\gamma$ & 1.9  &0.83  &  0.83 \\ 
$b$& 0.32  & 0.32 & 0.29 & $c$ & 0.72 & 0.74 &  0.74 \\
$\tau$& 0.65  &  0.60 &  0.60  & invis.& 0.21  & 0.21 & 0.21  \\  \hline\hline
\end{tabular}
\end{center}
\caption{Comparision of the results for Higgs
coupling uncertainties,  in \%,
 from data samples from the ILC combined with
those from LHC.  Each block of entries corresponds 
to an ILC stage.  For each entry, corresponding to the 
measurement of a Higgs coupling at that ILC stage,
the three columns represent:  first,  the 
entry in Table~\ref{tab:ILCstages}; second, the combination of this
data set with an LHC measurement of $BR(\gamma\gamma)/BR(ZZ^*)$
at 3000~fb$^{-1}$;
third, the combination of this data set with the results from the CMS
analysis for 3000~fb$^{-1}$ and Scenario 2, column CMS-HL-2 of 
Table~\ref{tab:CMSinterp}.}
\label{tab:ILCplusLHC}
\end{table}

\section{Editorial comments}

A number of aspects of this analysis deserve further comment:

\begin{enumerate}

\item  If we compare the ATLAS and CMS projections of Higgs rate
measurement accuracies side by side, as is done in
Table~\ref{tab:CMSinterp},
it is difficult not to conclude that the CMS projections are quite 
aggressive, even for Scenario 1.   One aspect of this comparison 
especially deserves comment.  

 There are three types of theoretical
uncertainties that contribute to the uncertainty in Higgs rates.  The first
is the uncertainty in the total cross section.  The second is the 
uncertainty in the probability of finding  a particular event property
used to search
for the Higgs events (for example, a jet veto). 
  The third is the modeling uncertainty involved
in determining the background in a signal region by extrapolation 
from a control region.  It is typical in LHC Higgs analyses that the 
Higgs contributes only 10\% of the total number of events in the 
signal region. The rest is SM background that must be 
subtracted.  To measure the Higgs rate to 5\%, it is 
necessary to normalize the background to 0.5\%.  It is 
often assumed that data from a control region determines
the background with precision that increases indefinitely with the 
statistics.  But, at some level, the uncertainties from
the model used for the extrapolation must be included.

In the CMS Scenario 2, only the first of these three types of 
theoretical uncertainty  is treated as an error that will be reduced by a
factor 1/2.  The other uncertainties are put into a category that is
decreased as $\sqrt{N}$, a factor of 1/11 between the current LHC
data set and the end of the HL-LHC running.  This prescription 
seems to overstate the value of the large statistics that the HL-LHC 
will acquire.

It is quite appropriate to take the CMS Scenario 2 estimates as a 
{\it goal} or an {\it opportunity} of the high-luminosity LHC stage.
But, for the reason just explained, 
I believe that it is not appropriate to treat these as an 
{\it expectation}.  Because of the dominance 
of systematic uncertainties, it is also not generally 
appropriate to combine 
projections from ATLAS and CMS.  My own judgement is
 that HL-LHC can expect to 
reach the CMS Scenario 1 accuracies. This will already add 
considerably to our knowledge of the Higgs boson.

\item   It is often commented that the LHC experiments can 
accurately measure ratios of Higgs branching ratios.  However,
detailed studies such as that in \cite{ATLASnew} find  large
theoretical errors for ratios of branching ratios.  This reflects
the different systems of cuts used to select events with 
different Higgs final states, which bring in theoretical errors 
from the 
second and third sources discussed above.  

Still, it is worth 
trying to create analyses in which theoretical errors cancel as 
much as possible. I believe that this is possible at least for the 
measurement of the ratio of branching ratios
$BR(\gamma\gamma)/BR(ZZ^*)$, as I have discussed in Section 6.
In  view of the key importance of this measurement for the long-term
program of Higgs coupling measurements, it would be valuable to 
define a precise protocol for extracting this quantity with minimal
systematic errors.

\item In contrast to the projections for LHC, the ILC projections are 
obtained as the result of full simulation in the environment in which 
the measurements are expected to be made~\cite{ILCHiggs}.  They are 
optimistic only 
in that the detectors described in the ILC TDR might be descoped as 
they move toward construction. On the other hand, some   properties estimated
for the TDR -- in particular, the efficiency for collecting
$\gamma\gamma$
events -- are known not to be fully optimized.  I believe that the 
ILC estimates {\it can} be treated as an expectation, and one that is 
likely to be surpassed with experience in operating the machine and
the 
detectors.

As is explained in \cite{ILCHiggs}, the ILC Higgs results depend on 
running the machine not only at 250~GeV but also at higher energies
where  $WW$ fusion production becomes important.   This is  
 necessary to provide enough statistics for the high-precision
 determination
 of the 
Higgs width, which determines the overall scale of partial widths.
The effect of this higher-energy running is seen most clearly in 
the left-hand panel of Fig.~\ref{fig:WZ}.   In contrast, the
$hZZ$
coupling, represented by the right-hand panel, is determined with 
high precision already at 250~GeV by the measurement of 
$\sigma(\ee \to hZ)$.    Precision determination of the Higgs 
width at 250~GeV is possible in principle, but it requires a 
multi-ab$^{-1}$ data set.

\item The treatment of the total
  Higgs boson  
width in making 
projections for $\ee$ colliders was controversial in the Snowmass 
Higgs study.  Blondel, in particular, argued eloquently for 
treating the Higgs total width as a free parameter, to be determined 
by the fit.  In his talk at the Seattle Energy Frontier meeting, he 
said that one should make ``no assumption on the Higgs 
exotic decays $\ldots$  thus making the fit truly model-independent
and truly representative of the lepton-collider
potential.''~\cite{BlondelSeattle}.

However, this approach is incorrect in that it 
does not take full account of the 
information that will be available from $\ee$ experiments,  especially
those with  tagged Higgs decays.  For example, the fits in 
\cite{Higgsworking} that use this prescription 
quote an uncertainty on the Higgs total width
of 5.0\% at the ILC500 stage.   The bulk of this uncertainty must
come from the presence of undetected exotic decay modes.
This should be compared to the results for the mode $h\to c\bar c$, 
which has
a branching ratio of 3\% in the SM and whose rate is 
expected to be measured
to 5\%  (0.15\% of the total width) at the same ILC stage.  The 
assumption in my fit is that undetected exotic decay modes 
have an upper limit of 0.9\% at this stage, similarly to the 
truly invisible modes.

In a tagged Higgs program, {\it all} decays of the Higgs
boson register experimentally in some way, so it is possible
to impose the constraint
\beq
          \sum_i BR(h\to i) = 1 \ .
\eeq{totalconst}
This constraint has a very powerful effect on the overall fit.
In Table~\ref{tab:global}, I compare the uncertainties on the 
Higgs total width obtained by the Snowmass Higgs Working 
Group, which did not apply the constraint \leqn{totalconst},
to those obtained from my fits.   I would claim that my fits
are equally model-independent to those in \cite{Higgsworking} 
but simply use more of the available information.

\begin{table}
\begin{center}
\begin{tabular}{l|c|c|c} 
ILC stage :   &  500  &   1000 & 1000up \\   \hline
Higgs width uncertainty, from \cite{Higgsworking} :  &   5.0   &   4.6 &   2.5  \\ 
Higgs width uncertainty, from this analysis :                   &
1.8  &  1.3  & 0.6     \\ 
\end{tabular}
\end{center}
\caption{Comparison of the uncertainty on the Higgs boson width, in 
\%, between the fits presented in the Snowmass Higgs working group report 
\cite{Higgsworking} and those given here.  }
\label{tab:global}
\end{table}

To make further progress in understanding the full power of 
precision Higgs measurements at the ILC, it would be interesting
to define a protocol that uses the power of the constraint
\leqn{totalconst} more directly.  An example of such an analysis
is the study of tau lepton decay branching ratios performed by the
Mark II experiment in the late 1980's~\cite{oneprong}.  
At the time, the sum of the measured 
branching ratios of the tau seemed to  deviate from 1, possibly
significantly.
This was called the ``tau 1-prong problem''.   The Mark II 
collaboration collected events in which one tau could be cleanly 
identified, then classified {\it all events} in this sample  into a set of categories
using the information 
from the opposite hemisphere.
   A key aspect of the analysis was 
that it defined  more categories than there are SM decay channels, so the
goodness of fit could test the hypothesis that there are no exotic 
tau decays~\cite{Burchat}.   The conclusion of the analysis, that 
small adjustments were needed in a number of measured branching
ratios, and that this eliminates the evidence for exotic tau decays, has stood up 
over the years.  We can take advantage of this strategy to design an
analysis that classifies all Higgs decay candidates at the ILC, so
that
the constraint \leqn{totalconst} can be applied with maximum power.
\end{enumerate}

\section{Conclusions}

The Snowmass 2013 study pointed out the importance of the precision 
measurement of the couplings of the newly discovered Higgs boson.
The study emphasized that the High-Luminosity LHC program will 
dramatically improve our knowledge of these couplings, and that further
qualitative improvements with important physics implications
are expected from measurements
at a lepton collider such as the ILC.    The analysis presented in
this  paper sharpens some of the points made in that study while 
reaffirming its general conclusions.  Now we patiently await the
data.

\bigskip

As I was completing this paper, I received a paper on the same 
subject by Han, Liu, and Sayre~\cite{Han}. That paper quotes
significantly larger uncertainties for the ILC projections.  These 
authors do not take into account most of the points made in 
Section 5 above.

\Acknowledgements

I am grateful to many people  who have helped me 
understand the methods used for estimation of 
Higgs boson coupling uncertainties in the Snowmass
study.  In particular, I have received especially 
valuable advice and criticism from 
Tim Barklow, Alain Blondel,  Patricia Burchat, 
Sally Dawson, Eric Feng,  Serguie Ganjour, Keisuke Fujii, Markus 
Klute, Tomohiko Tanabe, Jianping Tian, and Rick
Van Kooten.
This work was supported by the U.S. Department 
of Energy under contract DE--AC02--76SF00515.

\end{document}